\documentclass[reprint, superscriptaddress, amsmath,amssymb, aps, pra]{revtex4-1}

\usepackage{graphicx}
\usepackage{dcolumn}
\usepackage{bm}
\usepackage{float}
\usepackage{algorithm}
\usepackage{algpseudocode}
\usepackage{color}

\newcommand{\senstime}{\tau}
\newcommand{\realtime}{t}
\newcommand{\totaltime}{T}
\newcommand{\smalltime}{\senstime_0}
\newcommand{\maxpow}{K}
\newcommand{\curpow}{k}
\newcommand{\longtime}{\senstime_{\max}}
\newcommand{\freqvar}{\kappa}
\newcommand{\wien}{d\mathcal{W}}
\newcommand{\nn}{\nonumber \\}

\begin{document}

\title{Adaptive tracking of a time-varying field with a quantum sensor}

\author{Cristian Bonato}
\affiliation{%
Institute of Photonics and Quantum Sciences, SUPA, Heriot-Watt University, Edinburgh, United Kingdom}%
\author{Dominic W. Berry}
\affiliation{%
Department of Physics and Astronomy, Macquarie University, Sydney, NSW 2109, Australia}%

\begin{abstract}
Sensors based on single spins can enable magnetic field detection with very high sensitivity and spatial resolution. Previous work has concentrated on sensing of a constant magnetic field or a periodic signal.
Here, we instead investigate the problem of estimating a field with non-periodic variation described by a Wiener process. We propose and study, by numerical simulations, an adaptive tracking protocol based on Bayesian estimation. The tracking protocol updates the probability distribution for the magnetic field, based on measurement outcomes, and adapts the choice of sensing time and phase in real time. By taking the statistical properties of the signal into account, our protocol strongly reduces the required measurement time. This leads to a reduction of the error in the estimation of a time-varying signal by up to a factor 4 compared to protocols that do not take this information into account.
\end{abstract}

\maketitle

\section{Introduction}
Sensors based on individual quantum objects, such as electrons or atoms, can enable measurements of physical quantities with very high spatial resolution \cite{degen_arxiv_2016}.
Additionally, by exploiting quantum phenomena, one can reach a sensitivity beyond what possible with classical techniques \cite{GLM11}.
In the past decades, the exciting scientific progress in the control of  quantum systems has led to the demonstration of quantum sensing protocols based on individual photons, electrons, etc.
In this context, sensors based on single spins can map magnetic fields with nanometric spatial resolution, making them a revolutionary tool to study magnetic phenomena in nanoscale materials and biological processes \cite{schirhagl_review, wrachtrup_review_2016}. The most prominent system in this field is the electronic spin associated to the nitrogen-vacancy (NV) centre in diamond. Due to the weak spin-orbit coupling and an almost spin-free $^{12}$C environment, the NV centre spin preserves quantum coherence on timescales much longer than the manipulation time. Remarkably, the spin can be read out at ambient conditions by optically-detected magnetic resonance, making it a viable system for nanoscale magnetic sensing at ambient conditions.

These properties have led to ground-breaking experiments in nanoscale sensing, achieving a spatial resolution down to 10-20 nm \cite{wrachtrup_review_2016} and a sensitivity sufficient to detect individual electron spins \cite{grinolds_NNano2014} and nanoscale volumes of nuclear spins \cite{rugarNNano2015, haberleNNano2015, devienceNNano2015}, down to the individual nuclear spin level \cite{muller_NComms2014}. Remarkable experiments with NV centres include the application of nanoscale sensing to probe, for example, ballistic electron transport in a conductor \cite{kolkowitzScience2015}, topological magnetic defects \cite{dussauxNComms2016}, spin waves \cite{sarNComms2015} and vortices in superconducting materials \cite{thielNNano2016, pelliccioneNNano2016}.

The sensing capabilities of the NV electronic spin are not limited to magnetic fields but extend also to the measurement of other physical quantities such as temperature \cite{kucskoNature2013}, electric fields \cite{dolde_NPhys2011} and strain \cite{dohertyPRL2014}. Additionally, other defects in different materials, such as silicon carbide, exhibit sensing properties complementary to those of the NV centre in diamond \cite{falk_PRL2014, krausSiC_ScRep2014, niethammer2016, cochraneScRep2016}.

Quantum sensing experiments have mainly addressed the detection of constant (DC magnetometry) and periodic signals (AC magnetometry) \cite{taylorNPhys2008}. DC magnetometry estimates a constant signal by detecting its effect on a coherent superposition, e.g.\ by measuring a spin rotation under an applied constant magnetic field. AC magnetometry deals with detecting the amplitude and/or phase of a signal composed of one or a few harmonic tones by applying echo sequences. 

AC magnetometry was extended to the reconstruction of non-periodic waveforms \cite{cooperNComms2014} by using a family of echo sequences that form a basis for the signal. Identical instances of the same signal are repeated and detected using each echo sequence in the family, which allows the retrieval of specific Fourier coefficients. Combining all Fourier coefficients, corresponding to all the echo sequences in the family, the wavefunction can be reconstructed up to an arbitrary precision.
This waveform reconstruction technique enables the reconstruction of fast oscillating waveforms, but requires access to several identical instances of the same waveform in order to find the projection of the signal onto each echo sequence in the basis.

Here, we focus on a different problem: the reconstruction of a single instance of a time-varying magnetic field of known statistical properties. We propose a novel tracking protocol, based on Bayesian estimation, that extends a DC-magnetometry protocol to the estimation of a time-varying stochastic signal.  We study the protocol performance in the case of a Wiener process and show that our protocol reduces the estimation time by up to a factor 4 compared to known protocols in the literature, by taking the measurement history and the statistical properties of the signal into account.
A different method for the related problem of measuring a phase that changes in discrete steps was presented in Ref.\ \cite{wiebe_PRL_2016}.

Our protocol may find applications for fast tracking of magnetic fields associated to diffusion processes, for example in biology or in chemical reactions, or to track the Brownian motion of trapped magnetic nanoparticles. Additionally, this work could provide a faster way to track the dynamics of the spin bath surrounding the quantum sensor in the material. This could lead to a narrowing of the magnetic fluctuations and an increase of the spin coherence time \cite{cappellaroPRA2012, shulman_NComms2014}.

\section {Quantum sensing over a large dynamic range}
\subsection {Single-spin DC magnetometry}
\label{sec:intro}
In this subsection we summarize known techniques and results for measuring a constant frequency.
A constant magnetic field $B$, along the quantization axis $z$, can be measured by detecting the rotation induced on a single spin (Ramsey experiment) in the $xy$ plane. A spin initialized in the superposition state $(\left|0 \right \rangle + \left|1 \right \rangle)/\sqrt{2}$ evolves under $B$ over time $\senstime$ as $(\left|0 \right \rangle + e^{i \gamma B \senstime} \left|1 \right \rangle)/\sqrt{2}$, where $\gamma$ is the gyromagnetic ratio ($\gamma \sim 28$ MHz/mT for an electronic spin). Assuming perfect spin initialization and read-out and no decoherence, the probability to detect outcome $\mu \in\{0,1\}$ after time $\senstime$ is:
\begin{equation}
    p(\mu|f_B) = \frac{1 + (-1)^{\mu} \cos (2 \pi f_B \senstime + \theta)}{2}
\label{eq:ramsey}
\end{equation}
where $f_B = \gamma B/2\pi$. The phase $\theta$ corresponds to the rotation angle of the spin read-out basis in the $xy$ plane, relative to the initialization state. The goal of a sensing experiment is to retrieve the frequency $f_B$ with the highest possible accuracy over the largest possible range of values.

In realistic cases, the spin state associated with outcome $\mu$ can only be read out with finite fidelity $\xi_{\mu}$, defined as the probability to detect $\mu$ given that the eigenstate corresponding to $\mu$ is prepared. Additionally, the coherence of the spin is limited by fluctuations of the magnetic environment, averaged over the sensing time. We include magnetic fluctuations induced by nuclear spins in the material as a Gaussian decoherence term described by the coherence time $T_2^*$ \cite{delangeScience2010}. In our following discussion, we assume $T_2^*$ to be long ($T_2^* \sim 100~\mu$s) so that we can focus on reconstructing the variation of the classical magnetic signal neglecting the fluctuations of the nuclear spin environment. Coherence times of several hundred microseconds have been experimentally demonstrated in isotopically-purified diamond samples \cite{balasubramanian_NMat2009, bonatoNNano2016}.

Including finite read-out fidelity and decoherence, Eq.~\eqref{eq:ramsey} is modified to:
\begin{equation}
\begin{split}
    p(\mu=0| & f_B) = \frac{1+\xi_0 -\xi_1}{2} + \\
    & + \frac{\xi_0 + \xi_1-1}{2} e^{-\left ( \senstime/T_2^* \right )^2} \cos (2 \pi f_B \senstime + \theta)
\end{split}
\label{eq:ramsey_pk}
\end{equation}
and $p(\mu=1|f_B) = 1-p(\mu=0|f_B)$. In the following, based on our previous experiment with resonant optical excitation of an NV centre at cryogenic temperature \cite{bonatoNNano2016}, we assume good read-out fidelity for the outcome $\mu=1$ ($\xi_1 \sim 1$) and we only discuss the role of the read-out fidelity for outcome $\mu=0$, which we simply denote $\xi$.

One fundamental issue with frequency estimation by Ramsey measurements is the trade-off between sensitivity and measurement range. In other words, there is a limit on the dynamic range, defined as the ratio between the maximum measurable frequency before saturation ($f_{\rm max}$) and the smallest detectable frequency, described by the uncertainty $\sigma_f$.

For a Ramsey experiment with sensing time $\senstime$ repeated many times for total time $\totaltime$, the uncertainty $\sigma_f$ decreases as $1/(2\pi \sqrt{\senstime \totaltime})$. Therefore, the minimum uncertainty can be reached when measuring over the longest sensing time $\longtime$ before decoherence becomes significant, $\longtime \sim T_2^*$. On the other hand, the frequency range decreases with $\senstime$ because the signal is periodic, creating ambiguity whenever $\|2\pi f_{\rm max} \senstime\| > \pi$.
As a result, the dynamic range is bounded as $f_{\rm max}/\sigma_f < \pi \sqrt{\totaltime/\senstime}$.

Adaptive phase estimation protocols have been devised to overcome this limit.
The basic idea is to probe the field with a combination of $K+1$ exponentially decreasing sensing times $\senstime_{\curpow} = 2^{\curpow} \smalltime$, where $\smalltime$ is the smallest sensing time and $\curpow=\maxpow,\ldots,0$ \cite{higginsNature2007}.
In adaptive measurements, the phase $\theta$ is adjusted based on the results of previous measurements.
Provided there are multiple measurements for each sensing time, the uncertainty in estimating a phase scales as $1/N$, where $N$ is the total number of applications of the phase shift \cite{higginsNature2007}.

Further developments also showed that adaptive feedback is not a strict requirement: non-adaptive protocols can reach $1/N$ scaling \cite{higginsNJP2009, berryPRA2009, svore}.
In the case of frequency estimation, this translates to an increase in the dynamic range to $f_{\rm max}/\sigma_f \sim \pi (\totaltime/\senstime_0)$ \cite{hayesPRA2014, waldherr, nusran}.
For frequency estimation with realistic read-out fidelity, it was initially found that non-adaptive protocols yielded the best results \cite{saidPRB2011}, but later improvements were found using adaptive measurements \cite{hayesPRA2014,bonatoNNano2016}.

These protocols use Bayesian estimation.
The probability distribution $P(f_B)$ for the frequency $f_B$ is assumed to be uniform at the beginning of each estimation sequence,
then is updated after every Ramsey experiment according Bayes' theorem.
For the $\ell$-th Ramsey in the estimation sequence, Bayes' theorem gives
\begin{equation}
    P(f_B|\mu_1...\mu_\ell) \propto \color{black} P(f_B|\mu_1...\mu_{\ell-1}) P (\mu_\ell|f_B)
\end{equation}
where $P (\mu_\ell|f_B)$ is given by Eq.~\eqref{eq:ramsey}.
Although the frequency is not periodic, there are bounds to the range of possible values that will be considered, and the probability distribution for the frequency is periodic when multiples of $\smalltime$ are used for the sensing time.
It is therefore convenient to express the probability as a Fourier series
\begin{equation}
    P(f_B)=\sum_j p_j e^{i 2\pi j f_B \smalltime} \, .
\end{equation}
The coefficients $\{p_j\}$ depend on the measurement results, but that dependence will not be shown for brevity.

In the remainder of this subsection we summarize methods and results from Ref.~\cite{berryPRA2009}, except we replace the phase with $2\pi f_B \smalltime$.
When quantifying the performance of measurement of a periodic quantity, it is convenient to use the Holevo variance \cite{holevo1984}.
A modification of the Holevo variance, analogous to the mean-square error, is
\begin{equation}
    V_H := \left\langle \cos\left( 2\pi (f_B-\hat f_B) \smalltime \right)\right\rangle^{-2} -1 \, ,
\end{equation}
where $\hat f_B$ is the estimate of the frequency, and the average is over the actual frequency $f_B$ and measurement results.
This measure is convenient for designing the feedback protocol, but in this work we evaluate the performance of the estimation by the usual mean-square error.
The best estimate for the frequency, that minimizes $V_H$, is given by
\begin{equation}
    \hat f_B = \frac 1{2\pi \smalltime} \arg\left\langle e^{i2\pi f_B \smalltime} \right\rangle,
    \label{eq:avg_fb}
\end{equation}
where $f_B$ on the right-hand side is a dummy variable for the Bayesian phase distribution, and the expectation value is over that phase distribution.
This estimate is very easily found from the Fourier coefficients as
\begin{equation} \label{eq:est_fb}
    \hat f_B = \frac {1}{2\pi\smalltime}\arg \left| p_{-1} \right|.
\end{equation}

\begin{figure*}
\centering
\begin{picture}(600,100)
\put(0,-10){\includegraphics[scale=0.42]{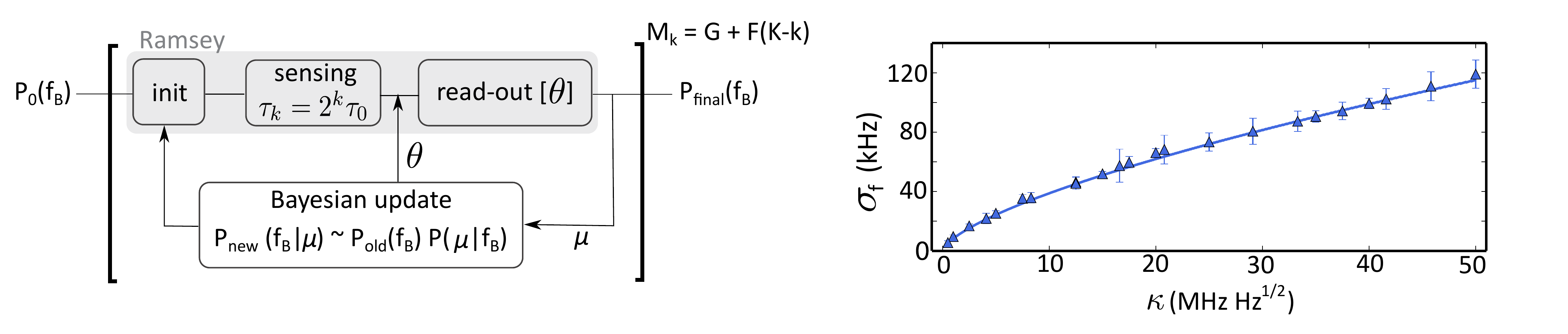}}
\put(0,90){(a)}
\put(290,90){(b)}
\end{picture}
\caption {(a) A conceptual diagram of the non-tracking adaptive estimation protocol \cite{bonatoNNano2016}. The protocol assumes an initial uniform distribution $P_0(f_B$). Each Ramsey experiment, comprising spin initialization, a sensing time $\tau_k$ and read-out, generates an outcome $\mu$ which is used to update $P(f_B)$ through Bayes' theorem. After each Ramsey experiment, the current distribution $P (f_B)$ is used to calculate the controlled phase $\theta$. Each Ramsey experiment, with sensing time $\senstime_{\curpow}$, $\curpow=\maxpow,\ldots,0$ is repeated $M_\curpow = G +(\maxpow-\curpow)F$ times, where $F$, $G$ are integers. At the end of the  sequence, the frequency $f_B$ is estimated from the final probability distribution $P_{\rm final}(f_B)$, according to Eq. \eqref{eq:avg_fb}. (b) The error $\sigma_f$ is plotted as a function of the fluctuation level $\freqvar$ [MHz Hz$^{1/2}$]. The points correspond to simulation results using the non-tracking protocol for $G=5, F=3$. The curve shows the prediction of Eq.\ \eqref{eq:eps_opt} with a fitted proportionality constant of $1.033$.}
\label{fig:eps_vs_dB}
\end{figure*}

In addition, the value of $\langle \cos( 2\pi (f_B-\hat f_B) \smalltime )\rangle$ in the expression for $V_H$ can be found by averaging over
$|\langle e^{i2\pi f_B \smalltime}\rangle|$.
Using the Fourier representation of the probability distribution, this means that $V_H$ is given by
\begin{equation}
    V_H = \left(2\pi \left\langle\left| p_1\right|\right\rangle\right)^{-2}-1,
\end{equation}
where the expectation is now over the actual frequency and measurement results, and we have used $p_1=p_{-1}^*$.
This suggests using an adaptive sensing protocol, where the rotation of the spin read-out basis $\theta$ is selected to maximize the expected value of $\left| p_1\right|$ after the next detection.
That minimizes the value of $V_H$ after the next detection.

The situation is more complicated, because initially large multiples of the interaction time $\smalltime$ are used, which means that $p_1=p_{-1}=0$.
When the smallest interaction time that has been used so far is $2^{\curpow}\smalltime$, then one would instead replace $\smalltime$ with $2^{\curpow}\smalltime$ in the above discussion.
In the approach of Ref.~\cite{higginsNature2007}, one would choose $\theta$ to maximize $\left| p_{2^{\curpow}}\right|$ after the next detection.
If the smallest interaction time so far was $2^{\curpow}\smalltime$, and a measurement to be performed is with an interaction time of $2^{\curpow-1}\smalltime$, then it is better to minimize $\left| p_{2^{\curpow-1}}\right|$ after the next detection.
The appropriate value of $\theta$ to choose is then
\begin{equation}
    \theta = \frac{1}{2} \arg \left( p_{-2^{\curpow}} \right).
\label{eq:theta}
\end{equation}

 The adaptive technique for realistic read-out fidelity in \cite{bonatoNNano2016} uses this estimate in combination with a phase increment  dependent only on the last measurement outcome, obtained by numerically optimizing the final variance for the specific experimental parameters through a “swarm optimization” procedure \cite{hayesPRA2014}.
In this work, we use that method for the measurements without tracking of the phase.
In contrast, for the measurements \emph{with} tracking of the phase, we always use Eq.~\eqref{eq:theta} to choose the controlled phase $\theta$.

\begin{figure}[ht]
\includegraphics[width=9cm]{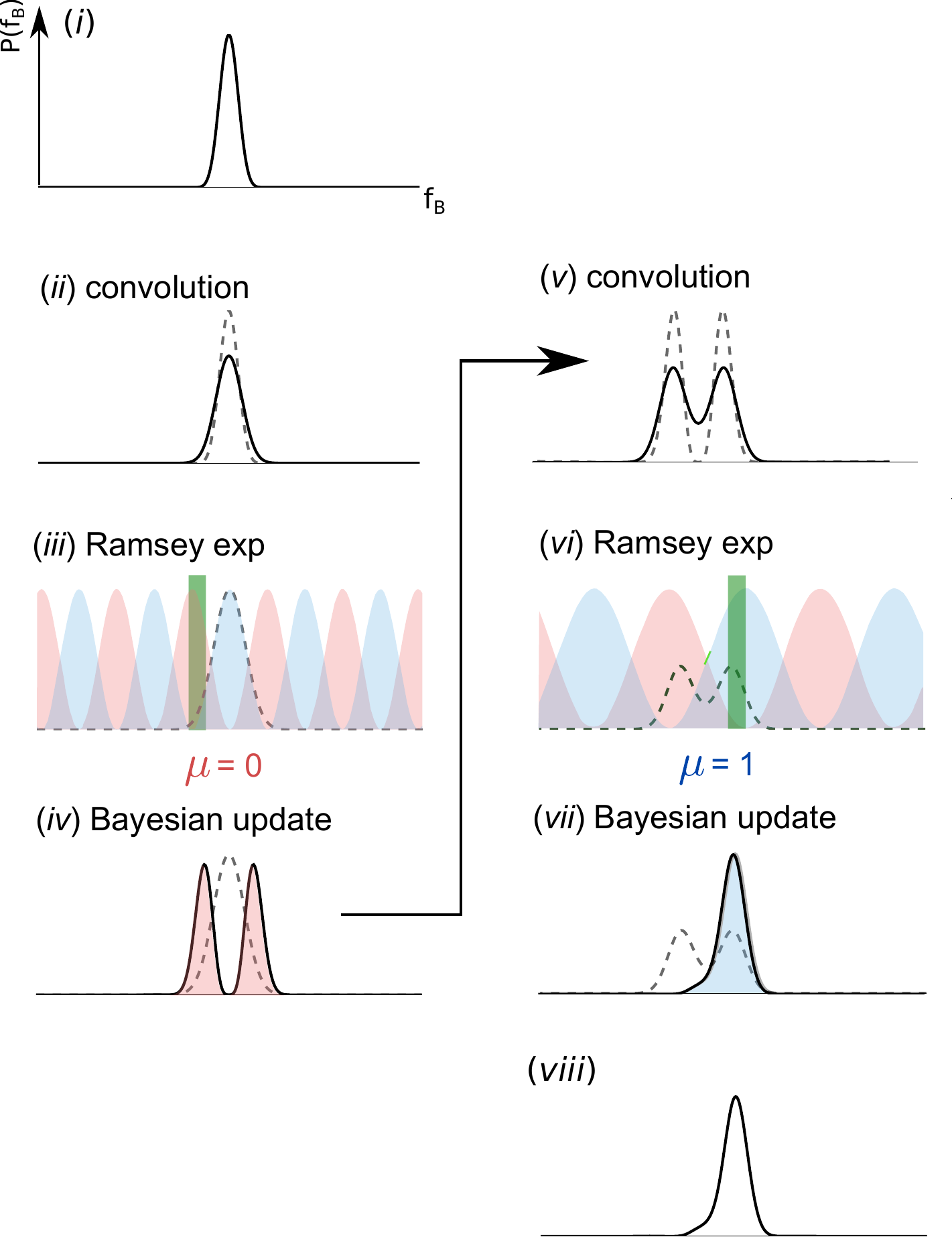}
\caption {Example of one estimation sequence for the tracking protocol described by the pseudo-code in Algorithm 1. The probability density for the current field estimation (\textit{i}) is convolved with the expected variation over the measurement time (\textit{ii}). A Ramsey experiment is performed: each of the two outcomes ($\mu = 0,1$) is associated to a conditional probability $P(\mu|f_B)$, represented, respectively, by red and blue shaded areas in (\textit{iii}). Outcome $\mu=0$ leads to the probability distribution $P'(f_B) \sim P_0(f_B)P(\mu=0|f_B)$ in (\textit{iv}). The figure of merit $\mathcal{F}$ for the two-peak distribution in (\textit{iv}) is below threshold and a new measurement is performed with a shorter sensing time, after a new convolution step (\textit{v}), to discriminate between the two peaks (\textit{vi}). Outcome $\mu=1$ leads to a narrow single-peaked distribution with figure of merit $\mathcal{F}$ above the threshold, which can be taken as a new estimate of $f_B$. The green vertical lines correspond to the actual values of $f_B$ at the specific time of each operation. Solid (dashed) black lines represent the current (previous) form of the probability distribution $P (f_B)$. }
\label{fig:pseudocode}
\end{figure}

In the following, we describe the estimation error for $f_B$ as the standard deviation based on the Holevo variance $V_H$ [$\sigma_f = V_H^{1/2}/(2\pi \tau_0$)]. While here we follow the traditional approach of giving a point estimate and the variance, an alternative possibility would be to give credible intervals for the estimate \cite{bayesian_book}, for example the frequency values $f_1$ and $f_2$ such that
\begin{equation}
    P(f_1< f_B< f_2| \mu_1...\mu_\ell) = 0.95.
\end{equation}
Alternatively, in our Bayesian approach the probability distribution $P(f_B)$ is available at all times after each measurement: neglecting memory and data processing considerations, giving $P(f_B)$, as shown  in Fig.~\ref{fig:pseudocode}(iv) and (vii), would provide the experimenter with the most complete information.

Estimation error scaling as $\sigma_f \propto 2^{-\maxpow}/\smalltime$ cannot be achieved by using only a single repetition for each of the exponentially-decreasing sensing times $2^{\curpow} \smalltime$.
However, the required bound can be reached using a number of repetitions $M_\curpow = G + (\maxpow-\curpow) F$,  where $F$, $G$ are integers \cite{higginsNJP2009,bonatoNNano2016}. For the longest sensing time $2^K \tau_0$, a number $G$ of repetitions is performed. The number of repetitions is then increased by $F$ for each shorter sensing time. The additional number of repetitions for shorter sensing times removes the most detrimental errors, which correspond to measurements that make the largest distinction in frequency $f_B$.
The total sensing time $T$ for a single estimation sequence is
\begin{equation}
    \totaltime = \smalltime [(2^{\maxpow+1}-1)G + (2^{\maxpow+1}-\maxpow-2)F]
\end{equation}
which can be approximated by
$\totaltime \sim (G+F) 2^{\maxpow+1} \smalltime$ for large $\maxpow$.

\subsection {Estimating a time-varying signal}
\label{sec:no-track}
Next we consider a frequency $f_B(t)$ varying according to a Wiener process
\begin{equation}
    f_B (\realtime + d\realtime) = f_B (\realtime) + \freqvar \, \wien(\realtime),
\label{eq:wiener}
\end{equation}
where $\wien(t)$ is an infinitesimal Wiener increment.
It is defined such that the integral of $\mathcal{W} (t)$ for time $\Delta \realtime$ has mean zero and variance $\Delta \realtime$.
It can be simulated by discretizing time to intervals of length $\Delta \realtime$, and generating a normal distribution with variance $\Delta \realtime$.
The goal is to estimate $f_B (t)$ in real time with the smallest possible error.
That is, we wish to estimate $f_B (t)$ at the current time using data up to the current time.

The simplest way to estimate a time-varying field is to repeat the optimized adaptive protocol described in Section \ref{sec:intro} and experimentally demonstrated in \cite{bonatoNNano2016}, which we will call the ``non-tracking protocol''. The phase acquired during a Ramsey experiment, which for a constant $f_B$ is simply $\varphi = 2\pi f_B \senstime$ [as in Eq.~\eqref{eq:ramsey}], becomes
\begin{equation}\label{phaseacc}
    \varphi = 2 \pi \int_{\realtime_0}^{\realtime_0+\senstime} f_B (\realtime) \, d\realtime \, .
\end{equation}
As discussed in Section \ref{sec:intro}, for a constant $f_B$ the minimum uncertainty is achieved by the longest sensing time $\longtime$ allowed by decoherence. In other words, the value of $\maxpow$ shall be chosen so that the longest sensing time $\longtime = 2^{\maxpow} \smalltime$ is close to the spin coherence time $T_2^*$.

This choice of $\longtime$ is not necessarily optimal for a time-varying signal. In this case, the optimal $\maxpow$ should be chosen such that the estimation error is similar to the change in signal over the measurement sequence (provided it is less than $T_2^*$). By choosing $\maxpow$ larger than this, the signal would fluctuate over the measurement time by more than the measurement accuracy, resulting in an unreliable outcome. On the other hand, if $\maxpow$ is smaller than necessary, then we would be restricting ourselves to a reduced accuracy. Therefore, we assume
\begin{equation}
    \sigma_f \propto \freqvar \, \totaltime^{1/2}
\label{eq:equality}
\end{equation}
where the total sensing time is $\totaltime \sim (G+F) 2^{\maxpow+1} \smalltime$.

For a measurement scheme of this type the uncertainty $\sigma_f$ should be inversely proportional to the total sensing time $\totaltime$.
More specifically, the scaling should be as $\sigma_f\propto 1/(2^\maxpow \smalltime)$, but the constant of proportionality will depend on $G$ and $F$.
A rough approximation may be made by assuming that the constant of proportionality is $1/\sqrt{G}$.
That is because the longest interaction time is repeated $G$ times, and $1/\sqrt{G}$ is the scaling for measurements repeated $G$ times.
The measurements with shorter interaction times have more repetitions, but are primarily used for resolving ambiguities.

That means we should have the scaling
\begin {equation}
2^{\maxpow} \smalltime \propto \frac 1{[G(G+F)]^{1/3} \freqvar^{2/3}}.
\end{equation}
That then yields an uncertainty scaling as
\begin {equation}
\sigma_{\rm no-tr} \propto \frac{(G+F)^{1/3}}{G^{1/6}} \, \freqvar^{2/3}.
\label {eq:eps_opt}
\end{equation}
The important part of this expression is the scaling with $\freqvar^{2/3}$, which is equivalent to the scaling that could be achieved for optical phase measurements with arbitrary squeezing \cite{BerryPRA02a,BerryPRA06a,BerryPRAE13}.
We have performed numerical simulations of this non-tracking protocol for $G=5$, $F=3$, and the results are shown in Fig.~\ref{fig:eps_vs_dB}.
There is excellent agreement with theory, and fitting for the proportionality constant for Eq.~\eqref{eq:eps_opt} yields $\sigma_{\rm no-tr} =  (1.033 \pm 0.04 ) \freqvar^{2/3}$.

In the previous discussion we assumed that the estimation sequence only includes sensing time, i.e.\ all other operations, such as spin initialization, control and read-out are instantaneous.
However, this is not true for realistic experiments, where all these operations contribute to an overhead time $T_{\rm OH}$.
A measurement sequence with $\maxpow+1$ different sensing times features a number of Ramsey experiments $R_\maxpow = (\maxpow+1)G + (\maxpow+1)\maxpow F/2$ \cite{bonatoNNano2016}, resulting in a total estimation time $\totaltime \sim (G+F) 2^{\maxpow+1} \smalltime + R_\maxpow T_{\rm OH}$.

\section {Adaptive tracking}
\subsection {The algorithm}
For a time-varying signal with known statistical properties, the available information can be exploited to shorten the estimation sequence.
Given a frequency $f_B(\realtime)$ at time $\realtime$, the frequency at time $\realtime + d\realtime$ is likely to be not too distant from $f_B(\realtime)$.
Using the known signal statistics, one can reasonably predict a narrower frequency range for the next estimation, so it is not necessary to explore the whole range of possible values.
In other words, instead of starting each estimation sequence from a uniform probability distribution, one could start from the Bayesian probability distribution from the prior measurements, and take into account the variation of the frequency according to the signal statistics.
This corresponds to tracking the time-varying signal.

\begin{figure}[h]
\begin{algorithm}[H]
\caption {Tracking protocol}
\begin{algorithmic}
\State $k = K$ (\textit{sensing time index, $\tau_k = 2^k \tau_0$})
\\
\While{TRUE}
\\
\State convolve ($\Delta t = 2^k \tau_0 + T_{\rm OH}$) - [\textit{Eq.~\eqref{eq:update_rule_convolution}}]
\State calculate $\theta$ - [\textit{Eq.~\eqref{eq:theta}}]
\State $\mu =$ Ramsey ($\tau_k = 2^k \tau_0$, $\theta$)
\State Bayesian update ($\mu$, $2^k \tau_0$, $\theta$) - [\textit{Eq.~\eqref{measup}}]
\State calculate $\mathcal{F}$ - [\textit{Eq.~\eqref{eq:fom}}]
\\
\State estimate $f_B$ - [\textit{Eq.~\eqref{eq:est_fb}}]
\\
\If {$\left ( \mathcal{F} < \mathcal{F}^{\rm (thr)}[k] \right)$}
\If {$\left ( k<K \right )$}
\State $k = k + 1$
\EndIf
\Else
\If {$k>0$}
\State $k = k - 1$
\EndIf
\EndIf
\EndWhile
\end{algorithmic}
\end{algorithm}
\end{figure}

\begin{figure*}[ht]
\centering
\centering
\begin{picture}(600,217)
\put(0,-10){\includegraphics[scale=0.38]{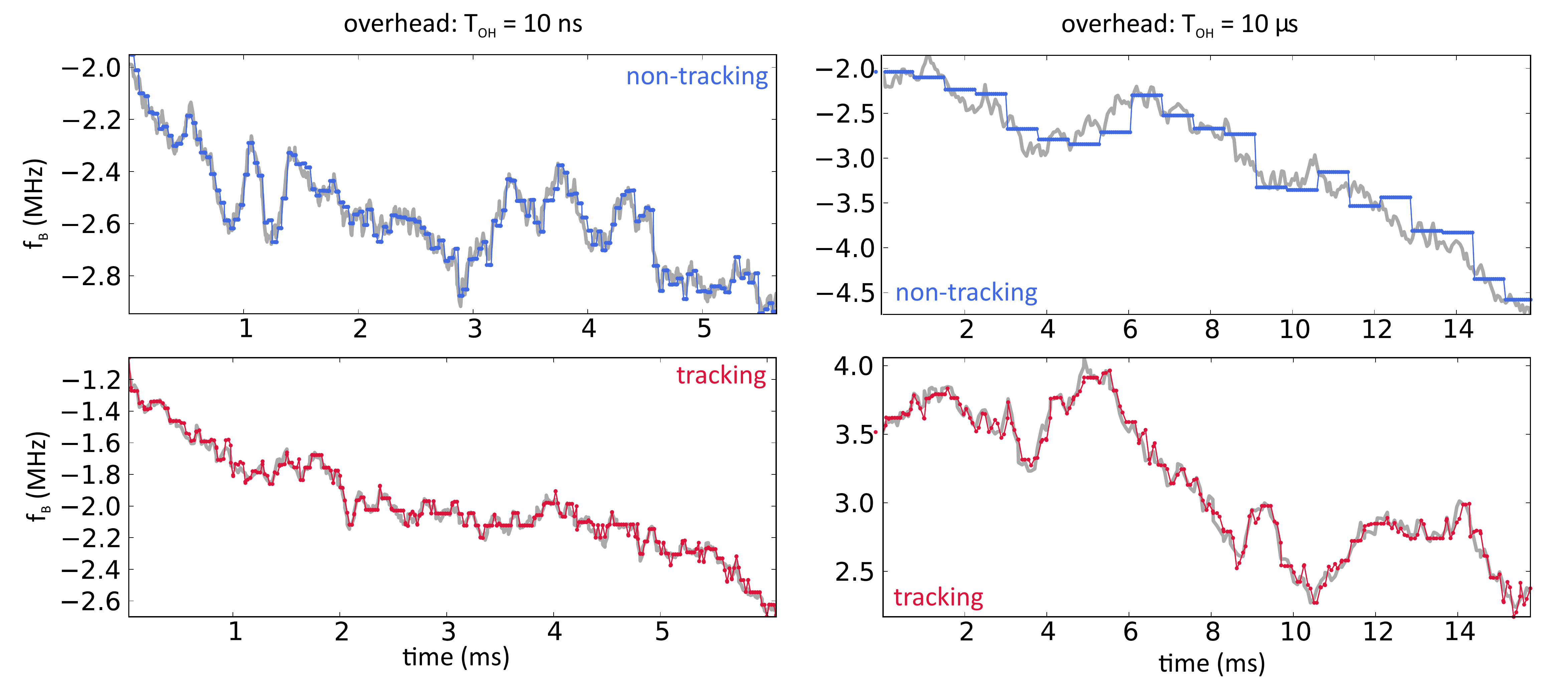}}
\put(0,205){(a)}
\put(260,205){(b)}
\put(0,95){(c)}
\put(260,95){(d)}
\end{picture}
\caption {Examples of waveforms (lighter-colour curves, in gray) reconstructed with the non-tracking [(a) and (b), darker (blue) curves] and tracking [(c) and (d), darker (red) curves] protocols.
On the left side, for (a) and (c), the overhead is small ($T_{\rm OH} = 10$~ns) and the measurement time mostly consists of the Ramsey sensing time.
In this case, the improvement by the tracking protocol is minimal.
On the right side, for (b) and (d), when the overhead due to spin initialization and read-out is more significant ($T_{\rm OH} =10~\mu$s), the advantage of the tracking protocol is clear.
In all plots, the time-varying signal $f_B (\realtime)$ (lighter gray curves) evolves according to a Wiener process described by Eq.~\eqref{eq:wiener}.}
\label{fig:waveforms}
\end{figure*}

\textbf {Updating the probability distribution}.
For simplicity, we will initially consider the case that the frequency is constant.
We will also take the fidelities $\xi_\mu$ to be equal to $1$.
Then given an outcome $\mu_\ell$ for the $\ell$-th Ramsey experiment, featuring a sensing time $\senstime_\curpow = 2^{\curpow} \smalltime$, the update of the probability distribution corresponds to an update of the Fourier coefficients as
\begin{align}\label{measup}
    p_j^{(\ell)} & = \frac{p_j^{(\ell-1)}}{2} + \frac{e^{-(\senstime_\curpow/T_2^*)^2}}{4}\left[ e^{i(\mu_\ell \pi +\theta)} p_{j-2^{\curpow}}^{(\ell-1)}\right.\nn
    & \quad \left. + e^{-i(\mu_\ell \pi +\theta)} p_{j+2^{\curpow}}^{(\ell-1)} \right].
\end{align}
Over a small time interval $\delta t$, the change in frequency $\delta f_B$ will have a normal distribution with variance $\freqvar^2 \delta \realtime$:
\begin{equation}
    P_G (\delta f_B) = \frac{1}{\sqrt{2\pi\delta t}\freqvar} e^{- (\delta f_B)^2/(2\freqvar^2 \delta t)}.
\label{eq:gaussian}
\end{equation}
Ignoring any information from a measurement, the probability distribution for the frequency after a time $\delta t$ will be the convolution of the initial probability distribution with the Gaussian in Eq.~\eqref{eq:gaussian}, giving
\begin{align}
    P^{(\ell)} (f_B) &= \int P^{(\ell-1)}(f_B - \nu) P_G (\nu)\, d\nu \nn
    &= \sum_{j} p_j e^{-2(\pi j \freqvar \smalltime)^2 \delta t} e^{i2\pi j f_B\smalltime}.
\end{align}
Therefore the coefficients $p_k^{(\ell)}$ for the probability distribution after a time $\delta t$ (without measurement) can be calculated as:
\begin{equation}
    p^{(\ell)}_j = p_j^{(\ell-1)} e^{-2(\pi j \freqvar \smalltime)^2 \delta t}.
\label {eq:update_rule_convolution}
\end{equation}
Provided the frequency does not vary significantly during an interaction time, the probability distribution may be approximated by using Eqs.~\eqref{measup} and \eqref{eq:update_rule_convolution} independently.
Using this approach the simulations still accurately model measurements made using this technique.
An exact calculation of the probability distribution could potentially result in more accurate estimates, but the method to perform such a calculation appears to be an open question.

\textbf {The adaptive tracking protocol}. The tracking protocol is described by the pseudo-code in Algorithm 1.
Each estimation sequence uses a probability distribution $P(f_B)$ based on the previous measurements.
As before, Ramsey experiments are performed starting from the longest sensing time $2^{\maxpow} \smalltime$, updating the probability distribution $P(f_B)$ according to Bayes' theorem.
The difference is that now, instead of using all sensing times $\tau_k$ with $G+(\maxpow-\curpow)F$ repetitions, the protocol adaptively chooses the best sensing time for each estimation, starting from the longest sensing time $2^{\maxpow} \smalltime$. To judge the accuracy of the estimate with sensing time $\tau_{\curpow}$, a figure of merit $\mathcal{F}$ is calculated based on the Bayesian probability distribution, and compared to a threshold $\mathcal{F}^{({\rm thr})}[\curpow]$. If the figure of merit satisfies $\mathcal{F} < \mathcal{F}^{({\rm thr})}[\curpow]$ then the same sensing time $\tau_{\curpow}$ is kept for the next estimation.
If the estimate is not sufficiently accurate, then the sensing time for the next estimation is reduced to $\tau_{\curpow-1}$. At that point, if the threshold is satisfied for the new estimation, the sensing time is increased back to $\tau_{\curpow}$. Otherwise, if the threshold is not satisfied, then the sensing time is further reduced to $\tau_{\curpow-2}$ and so on. While the non-tracking protocol requires a large number of Ramsey experiments for each estimation, the adaptive tracking protocol outputs an estimation of the time-varying field for each Ramsey experiment. 

We choose as a figure of merit $\mathcal{F}$ an estimate of the standard deviation of the probability distribution, which can be retrieved from the Holevo variance as:
\begin{equation} \label{eq:fom}
\mathcal {F} = \frac{V_H^{1/2}}{2\pi \tau_0} =  \frac{1}{2\pi \tau_0} \left [ \left(2\pi \left| p_{-1} \right|  \right)^{-2} - 1 \right]^{1/2}.
\end{equation}
This expression only depends on the coefficient $p_{-1}$ and can easily be calculated in real time. The estimation error is expected to scale as $\sigma_f \propto 2^K/\tau_0$. Therefore, we set the threshold corresponding to the sensing time $2^{\curpow} \tau_0$ to be:
\begin{equation}
 \mathcal{F}^{({\rm thr})}[\curpow] = \frac{\alpha}{2^\curpow\tau_0}.
\end{equation}
Numerical simulations suggest that the optimal value for the proportionality factor $\alpha$ is $0.15$.

This protocol is adaptive in two ways. First of all, it chooses in real-time the measurement phase according to Eq.~\eqref{eq:theta}. Second, it adapts the sensing time at each step, based on the current estimated variance. As a consequence, while the non-tracking protocol requires an optimal choice of $K$ for optimal performance (Section \ref{sec:no-track}) and sub-optimal choices of $K$ lead to large estimation errors, the tracking protocol is very robust and automatically selects the proper value of $k$ at each step.

Examples of reconstructions of time-varying fields with the tracking protocol, as compared with the non-tracking protocol described in Section \ref{sec:no-track}, are shown in Fig.~\ref{fig:waveforms}.  The non-tracking protocol can successfully estimate time-varying parameters, with no previous knowledge of the properties of the signal. When the overhead is small [Figs.~\ref{fig:waveforms}(a) and \ref{fig:waveforms}(c)], the performance of the non-tracking and tracking protocols appears to be similar. The advantage of the tracking protocol in following fast signal variations is clearly evident when overhead is large [Figs.~\ref{fig:waveforms}(b) and \ref{fig:waveforms}(d), for $T_{\rm OH} = 10~\mu$s].

\begin{figure*}[ht]
\centering
\centering
\centering
\begin{picture}(600,190)
\put(0,-10){\includegraphics[width=\textwidth]{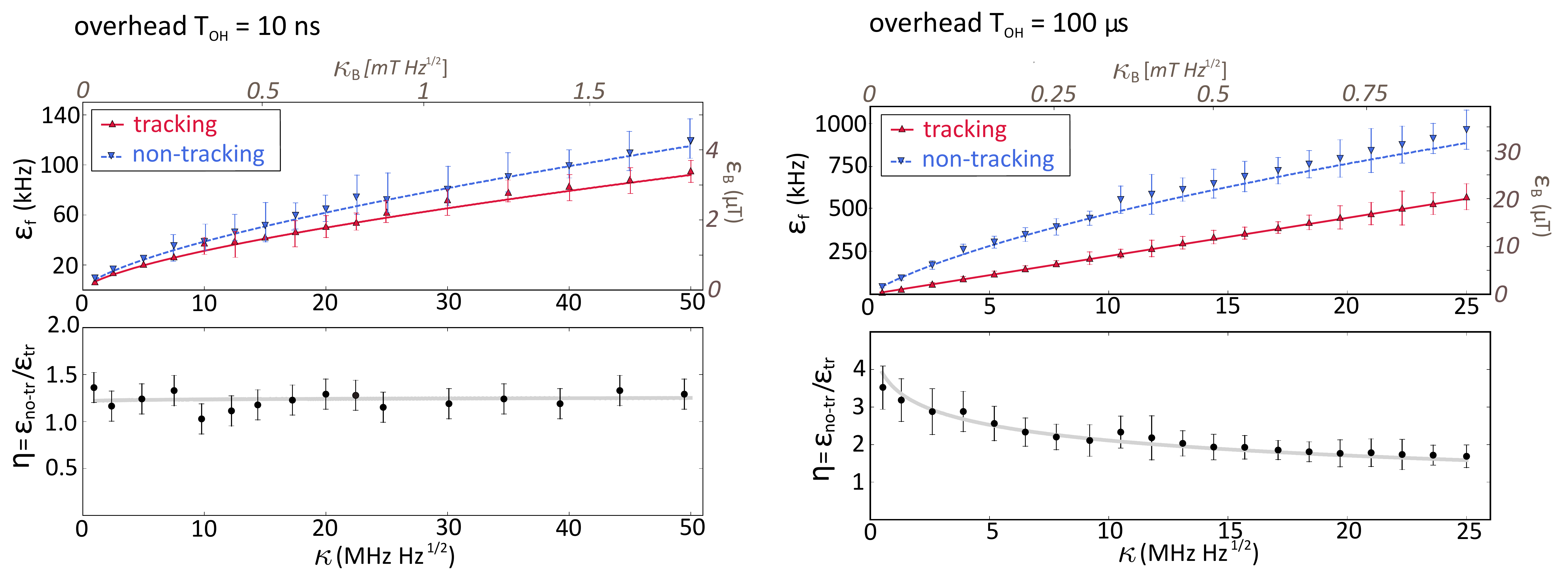}}
\put(0,175){(a)}
\put(260,175){(b)}

\end{picture}
\caption {Comparison between the performances of the non-tracking and tracking protocols in the limit of small overhead (a) and large overhead (b), for perfect read-out fidelity $\xi = 1$. In the upper graphs, the points show the waveform estimation error $\epsilon_f$ for the tracking (upward-pointing triangles, red) and non-tracking (downward-pointing triangles, blue) protocols, as a function of $\kappa$, from numerical simulations. The lines (solid for tracking, dashed for non-tracking) are the theoretical predictions based on Eq.~\eqref{eq:eps_opt} and Eq.~\eqref{eq:eps_opt_tr}, respectively, with the proportionality factors obtained by fitting. The grey axes on top and on the right show the same quantities specified for the case of a magnetic field sensor, representing $\kappa_B$ in [mT Hz$^{1/2}$] and the waveform estimation error in $\mu$T, respectively.  In the lower graphs, the ratio $\eta$ between the waveform estimation errors for the non-tracking and tracking cases is shown.}
\label{fig:sweepDB_fid100}
\end{figure*}

\subsection {Performance estimation}
\label{sec:perf_estim}
The shortest estimation sequence for the tracking protocol would be to simply repeat the longest sensing time $\longtime = 2^{\maxpow} \smalltime$ once, so that the estimation error scales as 
\begin{equation}
        \sigma_f \propto \frac {1}{ 2^{\maxpow} \smalltime}.
\end{equation}
Requiring that the variation of the frequency during a total sensing time of $2^K\tau_0$ is comparable to the uncertainty leads to the scaling
\begin{equation}
    \sigma_{\rm tr} \propto \freqvar^{2/3}.
\label{eq:eps_opt_tr}
\end{equation}
If we assume the scaling constant is $1$, the ratio $\eta$ between the errors in the non-tracking and tracking cases is:
\begin{equation}
    \eta = \frac{\sigma_{\rm no-tr}}{\sigma_{\rm tr}} = \frac{(G+F)^{1/3}}{G^{1/6}}.
\label{eq:eta}
\end{equation}
For $G=5$, $F=3$, $\eta \sim 1.5$.
The improvement expected for the tracking protocol is modest, and in simulations there is not a large difference,
as illustrated by the time-domain waveforms on the left side of Fig.~\ref{fig:waveforms}.
The reason for this is that the primary mechanism for improving the performance is using the prior information to resolve ambiguities, instead of measurements with shorter sensing times.
That means we can save the time used for the shorter measurements, and instead use the measurements with longer sensing times.
However, the contribution to the total time from the measurements with shorter interaction times is not large.
The longest sensing time $2^{\maxpow} \smalltime$ is almost the same as the sum of all the other sensing times $2^{\maxpow-1} \smalltime \ldots 2^0 \smalltime$, which means that the tracking protocol does not dramatically shorten the total estimation time.

A more consistent improvement can be expected when taking overhead into account. In the limit of large overhead, the major contribution to the estimation time is given by the overhead, while the sensing time can be neglected.
The estimation time for the tracking protocol can be approximated by $\totaltime_{\rm OH}$, as compared to $\totaltime \sim [(\maxpow+1)G + (\maxpow+1)\maxpow F/2] T_{\rm OH}$ for the non-tracking protocol. As an example, if $T_{\rm OH} =100~\mu$s, the tracking protocol delivers an estimation every $100~\mu$s. In contrast, the non-tracking protocol performs one estimation every $1.24$ ms (assuming $G=5, F=3, K=7$). Since the estimation error scales as the square root of the estimation time (Eq. \ref{eq:equality}), we expect an improvement on the order of $\eta \sim (1.24 \mbox{ms}/ 100 \mu\mbox{s})^{1/2}$, corresponding to a factor $3-4$.

\subsection {Numerical simulations}
The performance of the two protocols has been tested by numerical simulations, for a range of parameter values. We select the minimum sensing time $\smalltime = 20$ ns, corresponding to a frequency range $f_B \in [-25, +25]$ MHz.  An instance of a time-varying signal ${f}_B (\realtime)$ is produced according to the Wiener process in Eq.~\eqref{eq:wiener}, starting from a random value for $f_B(0)$, with a temporal resolution of $\smalltime = 20$ ns. In order to avoid values out of the  $[-25, +25]$ MHz range, the waveform is truncated if $|{f}_B|> 24$ MHz. The signal ${f}_B (t)$ is reconstructed using either protocol, providing the reconstructed waveform $\hat f_B(\realtime)$. To evaluate the performance of the estimation, we use the mean-square error
\begin{equation}
    \varepsilon^2 = \frac{1}{T} \int_0^T \left| {f}_B (\realtime) - \hat f_B(\realtime) \right|^2 d\realtime \, .
\end{equation}

Simulation results for the limit of negligible overhead ($T_{\rm OH} = 10$ ns) are shown in Fig.~\ref{fig:sweepDB_fid100} (a).
On the top plot, the tracking algorithm (blue downward-pointing triangles) exhibits a relatively small improvement compared to the non-tracking algorithm (red upward-pointing triangles), by a factor of $\eta = 1.23 \pm 0.09$, similar to the theoretical prediction $\eta \sim 1.5$ in the previous section. 
\begin{figure}[ht]
\centering
\includegraphics[width=9cm]{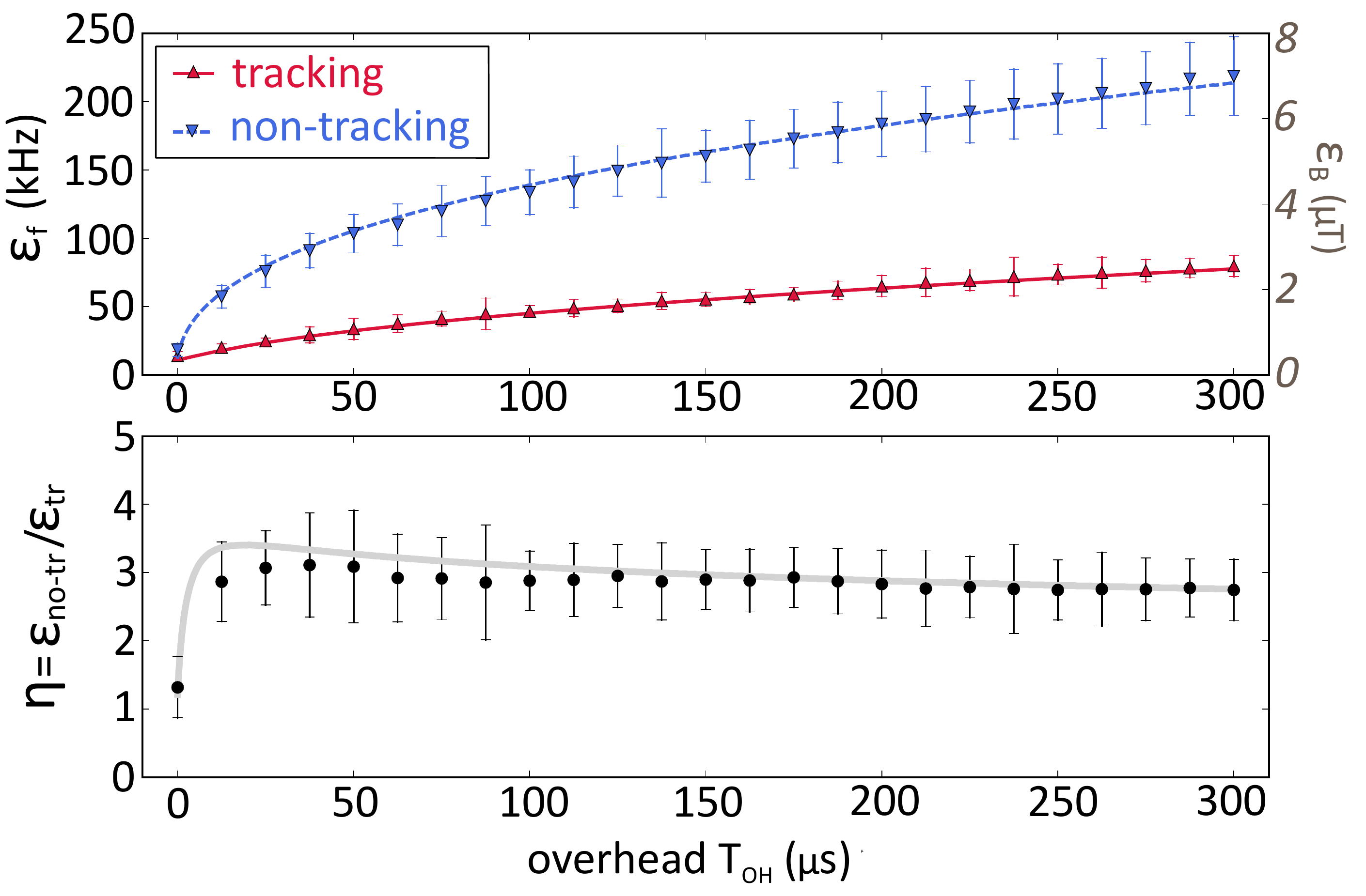}
\caption {Protocol performance as a function of the measurement overhead (qubit initialization and read-out time), for a fixed value for $\kappa = 2$ MHz Hz$^{1/2}$.}
\label{fig:ratio_vs_OH}
\end{figure}
The case of longer overhead is shown in 
Fig.~\ref{fig:sweepDB_fid100} (b). As hinted by the reconstructed waveforms in Fig.~\ref{fig:waveforms}, the advantage is here more significant, reaching $\eta \sim 3-4$.
Given the different approximations used, the theoretical predictions based on Eq.~\eqref{eq:equality} agree surprisingly well with the results of the numerical simulations.

The role of the overhead time is investigated in more detail by examining the protocols' performances for a fixed value of $\freqvar$ ($\freqvar = 2$ MHz Hz$^{1/2}$), while sweeping the overhead time $T_{\rm OH}$ between $0$ and $300~\mu$s. The results are plotted in Fig.~\ref{fig:ratio_vs_OH}. The improvement given by the tracking protocol (described by $\eta$) is small for small overhead, as already evidenced in Fig.~\ref{fig:sweepDB_fid100}. For larger overhead, $\eta$ is larger, up to about a factor $3$ and it is roughly independent of $T_{\rm OH}$, as predicted in Section \ref{sec:perf_estim}.

\begin{figure}[t]
\centering
\includegraphics[width=8.5cm]{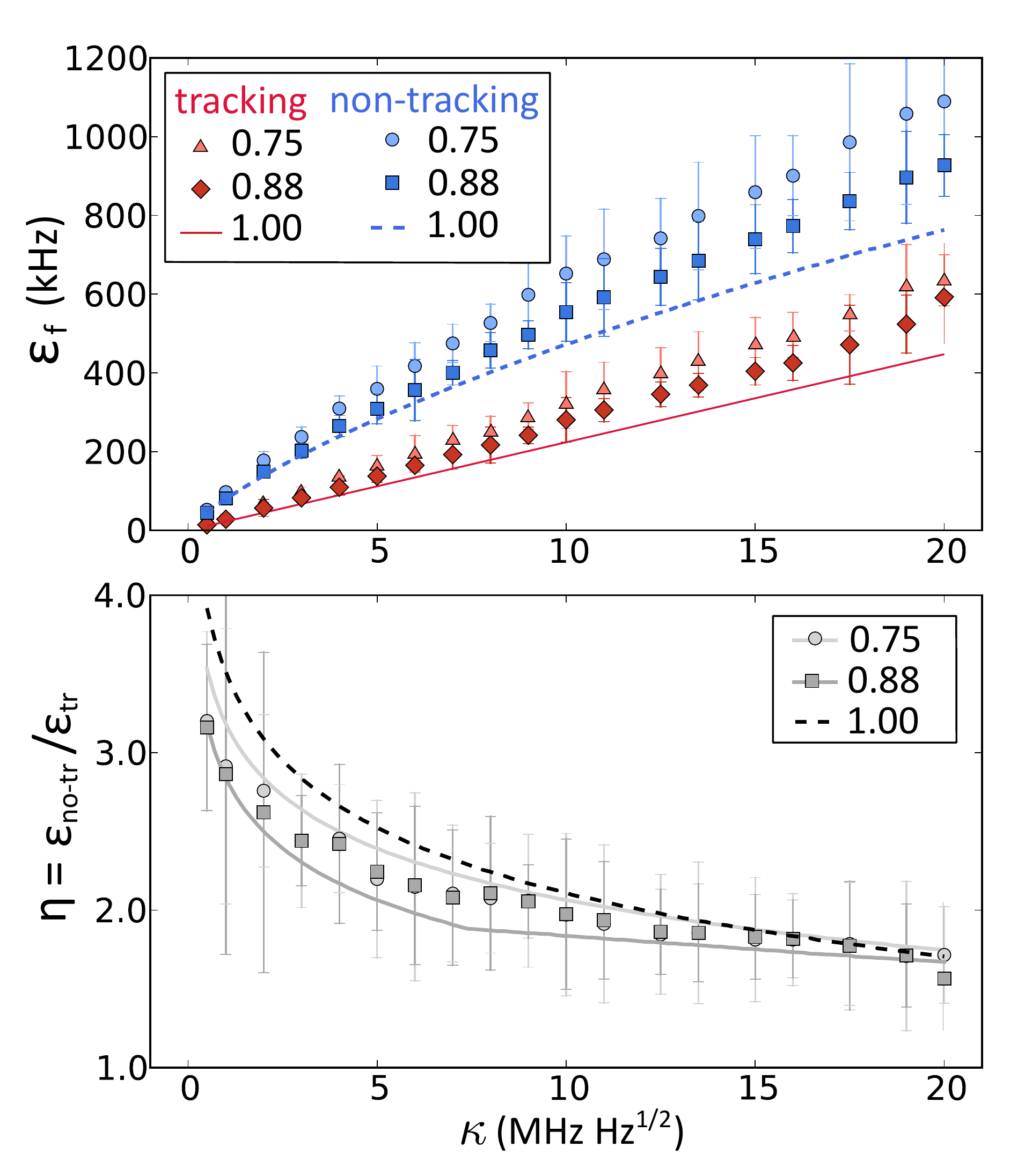}
\caption {Protocol performance for finite read-out fidelity $\xi$ in the non-tracking (blue points, square for $\xi=0.88$ and circle for $\xi = 0.75$) and tracking cases (red points, diamond for $\xi=0.88$ and triangle for $\xi = 0.75$). Lines (solid red line for tracking, dashed blue line for non-tracking) correspond to the estimation error in case of perfect fidelity ($\xi=1$). For all plots, the overhead is $T_{\rm OH} = 100 \mu$s. }
\label{fig:lower_fid}
\end{figure}

In the last set of numerical simulations (Fig.~\ref{fig:lower_fid}), we illustrate the effect of a reduced spin read-out fidelity $\xi$. We compare the waveform estimation error $\varepsilon_f$ for the non-tracking and tracking protocols for $\xi = 0.75$ and $\xi=0.88$. The latter value corresponds to the fidelity of spin read-out for the experimental demonstration in Ref.~\cite{bonatoNNano2016}. Reduced read-out fidelity leads, as one can expect, to an increase in the estimation error. The ratio $\eta$, however, does not vary significantly. 

\section {Experimental outlook}

Our analysis has been restricted to the case when the classical spin read-out noise is smaller than the spin projection noise, i.e.\ when ``single-shot'' read-out is available. Currently, single-shot read-out has only been experimentally demonstrated by exploiting resonant optical excitation of spin-dependent transitions at cryogenic temperature \cite{robledoNature2011}. This technique features high-fidelity spin initialization (fidelity $>0.99$) and  read-out (fidelities $\xi_1 \sim 0.99$ and $\xi_0 > 0.9$). The requirement of cryogenic operation is, however, a serious restriction for applications to quantum sensing.

Recent experiments have shown some promise towards the demonstration of single-shot read-out at room temperature. A first approach involves spin-to-charge conversion by optical ionization and successive detection of the defect charging state \cite{shieldsPRL2015, hopperPRB2016}. This approach may be further enhanced by the integration of electrical contacts to provide photo-electric spin read-out \cite{bourgeois}. A second approach involves the storage of the electron spin population onto the nuclear spin, which can be read out several times through the electron spin itself \cite{haberleRSI2017}. In both cases, a large overhead is introduced, which provides further motivation for our analysis.

Typically, the NV centre spin is read out at room temperature by spin-dependent photo-luminescence intensity, originating from spin-dependent decay rates through a metastable state under optical excitation (optically-detected magnetic resonance). In contrast to the ``single-shot'' readout case, the readout noise is here larger than the spin projection noise and spin read-out must be repeated several times for each sensing time. A large number of repetitions ($\sim 10^5$ for a typical experiment with NV centres) can be seen as a large overhead, suggesting that our tracking protocol will also be useful for such experiments.

In this work, we discussed the specific case of a time-varying signal described by a Wiener process, assuming that the parameter $\kappa$ is known. In case $\kappa$ is not known, one could simply start by measuring the signal evolution with the non-tracking protocol, and retrieving an estimate of $\kappa$ that can be used for tracking at a later stage. Additionally, our approach is quite general and can be easily extended to other kinds of stochastic processes. Particularly relevant is the Ornstein-Uhlenbeck process \cite{delangeScience2010}, which describes the fluctuations of a nuclear spin bath in the semi-classical approximation.

An interesting extension of this work would be to the case of the magnetic field induced by a quantum bath, such as the nuclear spin bath. If the correlation time of the bath dynamics is long enough, each estimation sequence provides a projective measurement of the magnetic field originating from the bath. The back-action of such projective measurement would narrow the probability distribution for the magnetic field induced by the bath, leading to an extension of the spin coherence time $T_2^*$ \cite{cappellaroPRA2012, shulman_NComms2014}. By providing faster field estimation, our protocol could allow to partially relax the requirement of long bath correlation time and slow dynamics. Additionally, spin read-out by optical excitation can induce perturbations in the bath by causing unwanted electron spin flips that affect the bath through the hyperfine interaction. By reducing the number of read-outs required for each estimation, a reduction of unwanted bath perturbations induced by optical spin read-out can be expected.

In addition to extending the sensor coherence time, the measurement back-action on the spin bath could also be used as a state preparation tool. For weakly-coupled nuclear spins in the bath, each Ramsey experiment in the estimation sequence acts as a weak spin measurement. It has been shown that a sequence of weak measurements with sensing times which are adapted in real time can be used as a tool for spin preparation \cite{blok_NPhys2014, greiner_purification_2016}. It would be interesting to investigate whether the tracking protocol could be adapted to provide, at the same time, the preparation of pre-determined quantum states of the bath.

A different research direction could be to extend this work to the spatial domain, to the field of microscopy. The acquisition of a bi-dimensional image requires considerable time, over which the system must be stable against fluctuations and drifts, posing a technological challenge. In this case, any a-priori knowledge of the statistical properties and correlation scales of the image could be used to speed up the measurement process \cite{singhPRA2015}.

\section{Conclusions}
In this work, we discussed the measurement of a single instance of a time-varying field, of known statistical properties, with a quantum sensor.  We investigated the performance of a non-tracking protocol, previously considered only for constant fields, which does not use any information about the signal. through numerical simulations, we showed that the protocol can successfully track a time-varying field. 
We introduced a novel tracking protocol based on Bayesian estimation. By using the additional information about the statistical properties of the signal, the tracking protocol shortens the time required for each estimation, leading to a reduced measurement uncertainty in the estimation of a time-varying field. While a small improvement is achieved when including only the sensing time, a considerable reduction in the estimation error, up to 4 times, is shown when taking the realistic measurement overhead (spin initialization and read-out time) into account.
Our findings can be relevant for fast tracking of time-varying magnetic fields associated to diffusion processes in biology and materials science. 

\section*{Acknowledgements}
The authors thank Machiel S.\ Blok, Hossein T.\ Dinani, Dale Scerri and Erik Gauger for helpful discussions. DWB is funded by an Australian Research Council Discovery Project (DP160102426).

%

\end{document}